\documentclass[10pt,journal,compsoc]{IEEEtran}

\usepackage{amsmath,amsfonts}
\usepackage{array}
\usepackage{textcomp}
\usepackage{url}
\usepackage{verbatim}
\usepackage{graphicx}

\usepackage[dvipsnames]{xcolor}
\usepackage[most]{tcolorbox}
\usepackage{multirow}
\usepackage{graphicx}

\usepackage{enumitem}
\usepackage{multirow}
\usepackage{algorithm}
\usepackage{algpseudocode}
\usepackage{booktabs} 

\newcommand{\ie}{\emph{i.e., }}
\newcommand{\eg}{\emph{e.g., }}

\newcommand{\header}[1]{\noindent$\bullet\quad$\textbf{#1.}}

\newcommand{\tabref}[1]{Table~\ref{#1}}
\newcommand{\figref}[1]{Figure~\ref{#1}}

\newcommand{\revise}[1]{\textcolor{black}{#1}}

\def\BibTeX{{\rm B\kern-.05em{\sc i\kern-.025em b}\kern-.08em
    T\kern-.1667em\lower.7ex\hbox{E}\kern-.125emX}}
\usepackage{balance}
\begin{document}
\title{ Reason4Rec: Deliberative User Preference Alignment of Large Language Models for Recommendation }

\author{Yi Fang, Wenjie Wang, Yang Zhang, Fengbin Zhu, Qifan Wang, Fuli Feng, and Xiangnan He
\thanks{This work was supported by the National Natural Science Foundation of China under Grants U25B2071 and 62572451, and by the Zhongguancun Academy under Grant C20250401. (\textit{Corresponding authors: Wenjie Wang and Fengbin Zhu.})}
\thanks{Yi Fang is with the University of Science and Technology of China, Hefei 230026, China, and with the Zhongguancun Academy, Beijing 100094, China (e-mail: peterfang@mail.ustc.edu.cn).}
\thanks{Wenjie Wang, Fuli Feng and Xiangnan He are with the University of Science and Technology of China, Hefei 230026, China (e-mail: wenjiewang96@gmail.com, fulifeng93@gmail.com, xiangnanhe@gmail.com).}
\thanks{Yang Zhang and Fengbin Zhu are with the National University of Singapore, Singapore 117417, Singapore (e-mail: zyang1580@gmail.com, zhfengbin@gmail.com).}
\thanks{Qifan Wang is with Meta AI, New York, NY 10001 USA (e-mail: wqfcr618@gmail.com).}
}


\markboth{Journal of \LaTeX\ Class Files,~Vol.~18, No.~9, September~2020}%
{How to Use the IEEEtran \LaTeX \ Templates}

\maketitle

\begin{abstract}

Aligning Large Language Models (LLMs) with recommendation tasks represents an emerging paradigm in recommendation domain, exhibiting promising performance overall. However, these aligned recommendation LLMs often struggle with complex scenarios due to limitations in the current alignment task formulation, which optimizes LLMs to directly generate user feedback without deliberation. To develop more reliable recommendation LLMs, we introduce a new \textit{Deliberative Recommendation} task, which enforces explicit reasoning about user preferences as an additional alignment objective. To address this task, we propose a \textit{Reasoning-powered Recommender} framework designed to enhance reasoning capabilities by leveraging verbalized user feedback in a step-wise manner. Specifically, this framework employs collaborative step-wise experts alongside specifically crafted expert-wise training strategies. Extensive experiments conducted on three real-world datasets demonstrate the rationality of the deliberative task formulation and the effectiveness of the proposed framework in improving both prediction accuracy and reasoning quality. Our implementation is publicly available on GitHub: \url{https://github.com/Peter-Fy/Reason4Rec}.


\end{abstract}


\section{INTRODUCTION}

%
%
%
%
%
Large Language Models are increasingly utilized to align with recommendation tasks through supervised fine-tuning~\cite{TallRec,Binllm,bigRec} or direct preference optimization~\cite{SDPO}. 
Although these recommendation LLMs (RecLLMs) generally achieve performance gains in different scenarios~\cite{DBLP:journals/www/WuZQWGSQZZLXC24, DBLP:conf/www/WangTHYLZZPW24, DBLP:journals/corr/abs-2410-19744}, they still fail in some complex cases, resulting in valueless and even disappointing recommendations. 
For instance, a RecLLM might recommend a vanilla‐attention tutorial to a senior Ph.D. who has already viewed LLaMA videos---relying solely on content relevance while disregarding that the senior Ph.D. already possesses the requisite knowledge.

Recent insights from OpenAI regarding the limitations of general LLMs~\cite{deliberative_alignment} suggest that failures of current RecLLMs come from their alignment objective, \ie optimizing the LLM to directly generate user feedback given the user history and candidate item. This pattern-based learning makes the model susceptible to data biases~\cite{DecodingMatters} and spurious correlations~\cite{yangzhang23,Xiangnan23}, hindering their ability to generalize to low-frequency or cold-start items. Moreover, this label-oriented optimization objective pushes the LLM to make predictions instantly, discouraging thorough reasoning about user preferences. Consequently, it is crucial to investigate alternative alignment objectives for RecLLMs. 

\begin{figure}[tbp]
\setlength{\abovecaptionskip}{0cm}
\setlength{\belowcaptionskip}{-0.5cm}
  \centering
  \includegraphics[width=\linewidth]{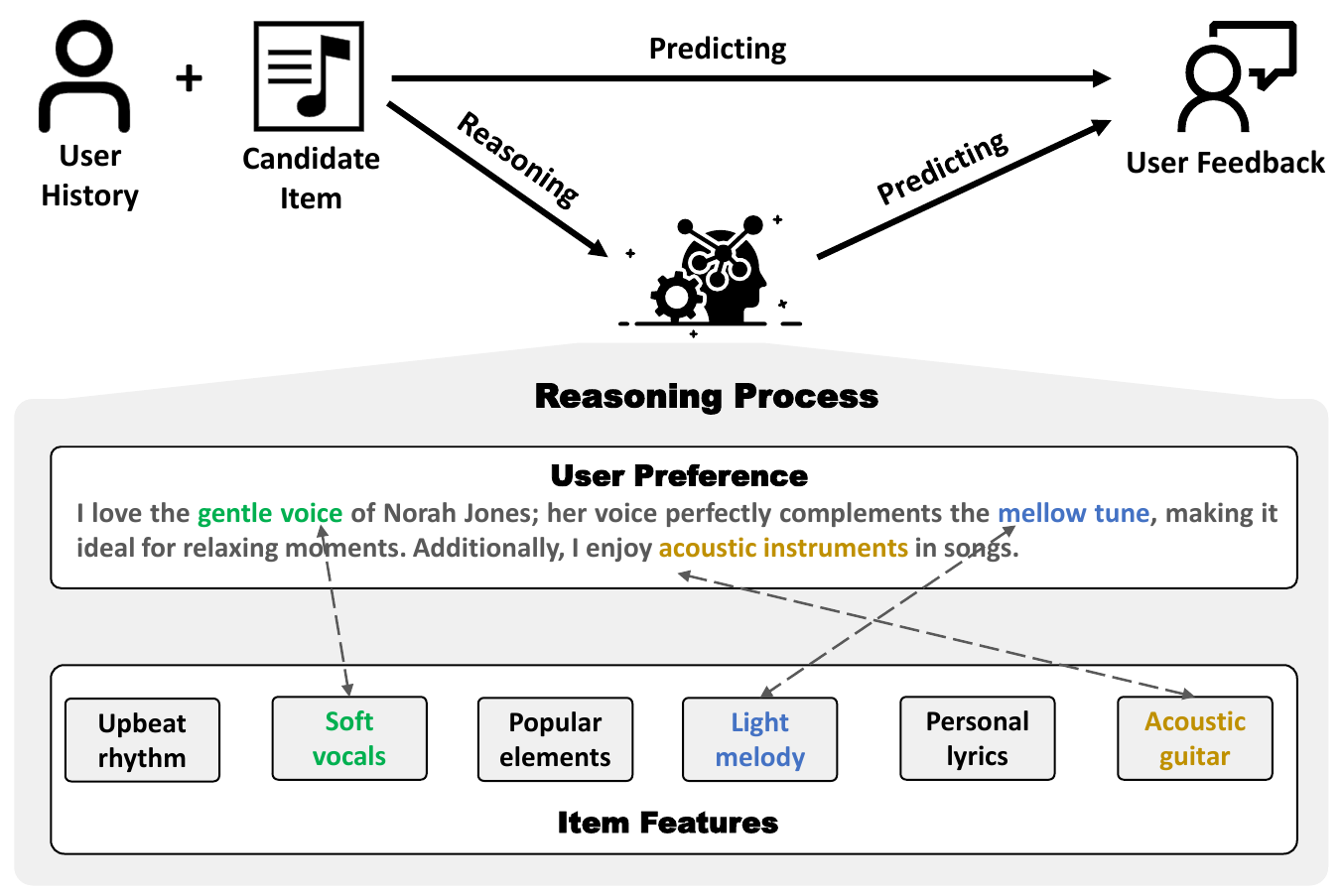}
  \caption{Comparison between the alignment objective of existing research, which optimizes LLMs to directly predict user feedback; and the objective of \textit{Deliberative Recommendation}, which optimizes LLMs to conduct explicit reasoning about user preferences before generating the prediction.}
  \label{fig:teaser}
\end{figure}

Inspired by the recent advancements in enhancing the reliability of general LLMs through explicit reasoning~\cite{cheng2025think,math1,math2,math3,logic1}, we introduce a new \textit{Deliberative Recommendation} task. This task is designed to train LLMs to engage in reasoning about user preferences and item features prior to predicting user feedback. As illustrated in Figure~\ref{fig:teaser}, the RecLLM will explicitly generate a reasoning process, using this process to augment its predictions. 
However, optimizing the RecLLM to generate explicit reasoning that aligns with user preference behind user feedback is quite challenging, since the user preference is highly personalized and diverse~\cite{Wang0WWF022}.
To address this challenge while adhering to commercial restrictions, the key of this formulation is seeking user generated and freely available rationale behind the user feedback as guidance for training the LLM in reasoning generation. Consequently, we incorporate verbalized user feedback, such as reviews or conversations, into our task formulation as supervisory signals for reasoning, setting our work apart from previous studies~\cite{ACL_findings,expert,GOT4Rec}.

To effectively address the Deliberative Recommendation task, it is crucial to decompose the highly personalized and diverse reasoning process into distinct steps, and equip the LLM with specialized reasoning capabilities for each step. In this way, we highlight three fundamental capabilities: 
1) \textit{Preference Distillation}, which analyzes the user history to identify aspect-level user preferences~\cite{guan2019attentive} and examines existing verbalized user feedback on the candidate item to recognize its positive and negative features; 
2) \textit{Preference Matching}, which matches distilled user preferences with candidate item features and generates rationales for why the user might like or dislike the candidate item; 
and 3) \textit{Feedback Prediction}, which predicts the user feedback with consideration of the generated rationales.


In this light, we propose a \textit{Reasoning-powered Recommender} (Reason4Rec) framework to teach LLMs with the recognized reasoning capabilities, guided by verbalized user feedback (\eg user reviews~\cite{ReviewRecSurvey}). As illustrated in Figure~\ref{fig:framework}, Reason4Rec employs 
three collaborative experts, each dedicated to a specific reasoning step, to hone step-specific reasoning capabilities. 
To reduce the computational and memory cost, these experts are implemented using separate QLoRA~\cite{Qlora} adapters on the same LLM. 
Recognizing that each reasoning step necessitates extracting distinct signals from the verbalized user feedback, each expert is equipped with a uniquely tailored training strategy. 
We conduct extensive experiments across three real-world datasets, validating the effectiveness of Reason4Rec regarding prediction accuracy and reasoning quality, and the rationality of the framework design. 
%




%
%
%
%
%
To summarize, our contributions are threefold:
\begin{itemize}[leftmargin=*]
\item We formulate the \textit{Deliberative Recommendation} task, in which LLMs engage in explicit user preference reasoning before making feedback predictions. To align the reasoning with the user's true preference, we propose to adopt verbalized user feedback to guide the learning process.

\item We propose a \textit{Reasoning-powered Recommender} framework for deliberative user preference alignment, which achieves the reasoning process with step-wise experts associated with specifically designed training strategies. 

\item We conduct experiments on three datasets, validating the effectiveness and rationality of the proposed Reason4Rec framework, showing the potential of slow thinking in recommendation.

\end{itemize}

\section{RELATED WORK}
In this work, we focus on Deliberative Recommendation, empowering RecLLMs with explicit reasoning abilities and optimizing LLMs through verbalized user feedback. This is highly related to LLM-based recommendation, deliberative alignment, and review-based recommendation. 

\subsection{LLM-based Recommendation}
Recent research has shown growing interest in RecLLMs~\cite{DBLP:journals/corr/abs-2410-19744}, with many studies exploring the use of LLMs to predict user feedback through in-context learning~\cite{chatgpt_CIKM,DBLP:conf/recsys/DaiSZYSXS0X23,DBLP:conf/recsys/SannerBRWD23} and fine-tuning~\cite{TallRec,Binllm,collm}. 
However, these approaches often instruct LLMs to generate predictions directly without disclosing their intermediate reasoning steps, 
limiting the utilization of LLMs’ reasoning capabilities.
To address this limitation, some recent studies explored leveraging LLMs' reasoning capabilities for recommendation tasks through various prompting strategies, such as chain-of-thought (CoT)~\cite{CoT} and self-reflection~\cite{DRDT}.
For instance, DRDT~\cite{DRDT} encourages LLMs to perform sequential recommendation in a divergent thinking manner by prompting them to analyze user preferences from multiple aspects, while simultaneously reflecting on their analysis using a critic prompt~\cite{DRDT}. 
GOT4Rec~\cite{GOT4Rec} applies the graph of thought strategy to prompt LLMs to reason through three different directions.
However, these in-context learning methods are inherently constrained by the models' existing capabilities, making LLMs often struggle to handle recommendation tasks effectively without task-specific fine-tuning~\cite{TallRec}.

\begin{figure}[tbp]
\setlength{\abovecaptionskip}{0cm}
\setlength{\belowcaptionskip}{-0.5cm}
  \centering
  \includegraphics[width=\linewidth]{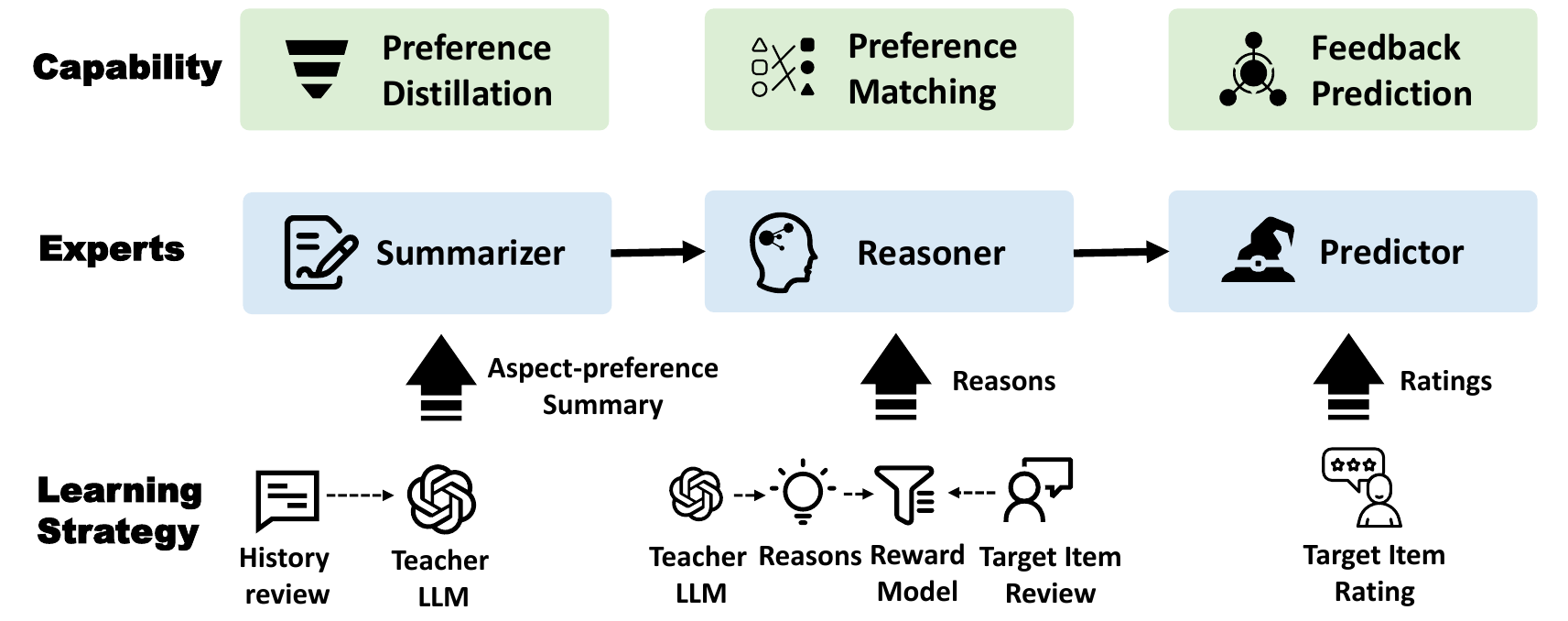}
  \caption{Illustration of the Reasoning-powered Recommender framework.}
  \label{fig:framework}
\end{figure}

Some studies have explored fine-tuning LLMs to improve reasoning ability specifically for RecLLMs~\cite{expert,ACL_findings,ReasoingRec,RecReasonor}. 
For example, RecSAVER utilizes reasoning generated by a larger LLM as ground truth to fine-tune a smaller LLM, enabling the latter to learn reasoning patterns tailored for recommendation tasks~\cite{ACL_findings}.
EXP3RT~\cite{expert} employs a similar fine-tuning strategy but introduces an additional preparatory step: constructing user and item profiles based on their review histories first. 
Leveraging these constructed profiles, EXP3RT reasons about user preferences and predicts possible ratings in the same single step.
However, these fine-tuning methods face two main limitations. First, they compress the complex reasoning process into a single step and train it jointly, which makes it challenging for LLMs to effectively learn. 
In contrast, Reason4Rec decomposes the reasoning process into multiple steps, training each step independently for better performance.
Second, they lack fine-grained supervision in training data selection, leading models to learn the reasoning patterns misaligned with users' real preferences. 
Reason4Rec addresses this limitation by using verbalized user feedback as supervision, resulting in reasons that more accurately reflect users’ preferences and enhance the multi-step reasoning abilities of RecLLMs.

\subsection{Deliberative Alignment}

Existing work has demonstrated the superiority of LLMs combined with multi-step reasoning~\cite{wang2024q}, with OpenAI's O1 showcasing remarkable reasoning capabilities~\cite{openai-o1}. 
Meanwhile, Deliberative Alignment~\cite{guan2024deliberative}, a training paradigm that directly teaches LLMs human-written safety specifications and trains them to explicitly reason about these specifications before responding, has garnered widespread attention in the LLM safety domain.

However, there are significant differences between this and the Deliberative Recommendation task: 
1) in the safety domain, alignment specifications are relatively well-defined and consistent, whereas the reasoning process behind user feedback is highly complex, with significant personalization and diverse preferences. 
Achieving deliberative user preference alignment at the individual level is thus a far more challenging task. 
2) Given the complexity of the Deliberative Recommendation task, relying solely on numerical user feedback (\eg ratings) or pairwise preference data for LLM learning, is insufficient. We propose leveraging verbalized user feedback to supervise reasoning for superior alignment, setting it apart from previous approaches to Deliberative Alignment.

\subsection{Review-based Recommendation}
Review-based recommendation systems have been widely studied recently to address tasks such as rating prediction~\cite{rating_predict1, rating_predict2,rating_predict3,rating_prediction4,rating_prediction5}. 
These methods typically adopt a dual tower architecture that employs a user encoder and an item encoder to capture the semantics of reviews as user and item embeddings. 
Then, the user and item embeddings are combined through a fusion layer to derive joint representations, which are passed to a rating prediction layer to produce the final output~\cite{ReviewRecSurvey}.
To obtain better embeddings of users and items from reviews, many advanced techniques have been explored. 
For instance, DeepCoNN~\cite{deepconn} employs convolutional neural networks (CNNs) to extract semantic information from historical reviews, achieving excellent results. Furthermore, attention mechanisms~\cite{attention} are also introduced at both the word and review levels to identify critical words and reviews that are most significant~\cite{liu2019nrpa,attentionRec2,attention3}. 
For example, NARRE~\cite{NARRE} employs attention mechanisms to prioritize reviews that are most useful for improving embedding learning.
While these methods leverage reviews to represent user preferences implicitly with embeddings, resulting in the lack of interoperability. 
In contrast, Reason4Rec generates the reasoning process for each step in explicit textual form, offering clearer and more interpretable explanations of user preferences. 
Additionally, most review-based recommendation methods primarily leverage reviews as input features for generating embeddings. However, Reason4Rec also utilizes reviews as supervisory signals on the output side, aligning the LLM’s reasoning with the user's true preferences.

\section{TASK FORMULATION}

\noindent$\bullet$ \textbf{Deliberative Recommendation.} 
We formulate the task of deliberative recommendation, which performs thoughtful reasoning regarding user preference and item features before predicting user feedback as illustrated in Figure~\ref{fig:teaser}. 
To guide the reasoning process, we propose incorporating verbalized user feedback for LLM optimization. 
In this work, we adopt user reviews due to fine-grained user preferences described in reviews~\cite{ReviewRecSurvey}. 
Besides, we investigate deliberative recommendation on the classic rating prediction task, following prior work~\cite{ACL_findings,expert}.

Formally, let $\mathcal{U}$ denote a set of users and $\mathcal{I}$ a set of items. 
The interaction between user $u$ and item $i$ can be represented as $(r_{ui}, c_{ui})$, where $r_{ui}$ is the rating of user $u$ for item $i$, and $c_{ui}$ is the corresponding review.  Such paired data is widely available in many datasets~\cite{movieLen,KaggleMovie}. 
For a user-item pair $(u, i)$, the historical interactions of user $u$ and item $i$ are defined as:  
$$
\begin{cases}
    \mathcal{H}_u = \{(i, r_{ui}, c_{ui}) \mid i \in \mathcal{I}_u\}, \\
    \mathcal{H}_i = \{(u, r_{ui}, c_{ui}) \mid u \in \mathcal{U}_i\},
\end{cases}
$$
where $\mathcal{I}_u$ denotes items previously rated by user $u$, and $\mathcal{U}_i$ denotes users who rated item $i$. 
Given $\mathcal{H}_u$ and $\mathcal{H}_i$, the LLM predicts the rating $\hat{r}_{ui}$ after performing explicit reasoning, where $\text{reason}_{ui}$ denotes the reasoning process and $\hat{r}_{ui}$ is the predicted rating:
\[
\text{reason}_{ui}, \hat{r}_{ui} = \text{LLM}(\mathcal{H}_u, \mathcal{H}_i).
\]

Considering the reasoning process is highly personalized and diverse across users, we decompose the above reasoning and prediction process into multiple steps, as shown in Figure~\ref{fig:framework}, to achieve different objectives: Preference Distillation, Preference Matching, and Feedback Prediction. 
To emphasize the alignment between the reasoning process and the user's true preference, we incorporate the target item review $c_{ui}$ of user $u$ for item $i$ as additional supervision during training. 
Notably, it is excluded in testing to prevent data leakage. 

\noindent$\bullet\quad$\textbf{Justification for explicit reasoning in RecLLMs.} 
Incorporating explicit reasoning into RecLLMs holds significant potential, despite it potentially increases the prediction time and computational cost. 
Specifically, it enables RecLLMs to: 1) fully leverage the reasoning capabilities of LLMs to improve prediction accuracy and reliability; 
2) enhance interpretability through step-by-step reasoning as explanations, thereby enhancing user trust; 
3) possibly facilitate user-controllable recommendation for human-AI collaboration~\cite{yang-etal-2024-human}, since users can adjust the reasoning process to generate new recommendations. 
Notably, explicit reasoning is particularly beneficial in scenarios where real-time recommendation requirements are less critical, but accuracy is paramount, such as E-commerce product, travel plan, and medication recommendations~\cite{drugRec}. 
These items typically involve long usage periods and require careful deliberation before purchase or viewing. 
Furthermore, as LLMs efficiency is continually improving~\cite{data_efficient}, the computational costs will decrease accordingly, enhancing the feasibility of wide applications. 
\section{METHOD}



In this section, we present Reason4Rec with three experts to collaboratively perform multi-step reasoning before recommendation. 
To supervise the reasoning process, Reason4Rec incorporates user reviews, directly aligning the reasoning with user preferences rather than only inferring preferences through users' numerical or binary feedback (\eg ratings and clicks). 
We then detail the three experts, \ie summarizer, reasoner, and predictor, with their fine-tuning strategies. An overview of Reason4Rec is illustrated in Figure~\ref{fig:framework}. 





\subsection{Summarizer for Preference Distillation} \label{sec:step1}


Reason4Rec leverages a summarizer to distill the user's preferences and the item's positive and negative features from their historical reviews. 
It filters out noise from the reviews, captures nuanced user preferences, and enriches item features with the LLM's inherent knowledge. 
The distillation output is formatted as an aspect-preference summary, 
\ie concise and representative keywords that encapsulate essential user preferences and item features, providing a reasoning foundation for the subsequent reasoning process. 


\noindent$\bullet\quad$\textbf{Summarizer Instantiation.} 
Formally, given a historical interaction $(r_{ui}, c_{ui})$ between user $u$ and item $i$, the summarizer will generate an aspect-preference summary. 
The summarizer's prompt is illustrated in the following, which uses the Music dataset~\cite{amazon_product} as an example. 

\begin{figure}[h]
\begin{center}
\fcolorbox{black}{gray!6}{\parbox{0.98\linewidth}{\centering 
\textbf{Summarizer Prompt} \\[2pt]
\raggedright 
\textbf{Task:} Summarize the reasons behind the given rating of a Music based on the customer review. \text{\textbackslash n} \textbf{Music:} $i$ \text{\textbackslash n} \textbf{Rating:} $r_{ui}$  \text{\textbackslash n} \textbf{Review:} $c_{ui}$  \\[2pt]
Analyze the customer \revise{$u$'s} review for Music $i$ and summarize the reasons behind the given rating of $r_{ui}$. Please consider the positive and negative aspects mentioned in the review and provide the keywords of reasons and user preference elements. 
}} 
\end{center}
\end{figure}

The user's preferences and the item's features are outputted as an aspect-preference summary. A template example is as follows: 
\begin{center}
\fcolorbox{black}{gray!6}{\parbox{0.98\linewidth}{
\textbf{Positive Aspects:} Catchy Melody, Unique Instrumentation, ... \\
\textbf{Negative Aspects:} Repetitive Lyrics, Overuse of Autotune, ...  \\
\textbf{User Preference Elements:} Harmony, Emotional Resonance, ... 
}} 
\end{center}

\noindent$\bullet\quad$\textbf{Summarizer Fine-tuning.} 
Considering ChatGPT~\cite{chatgpt} has shown strong performance in various summarization tasks~\cite{textSummary, text_summarization, text_summarization2}, we employ closed-source ChatGPT as a teacher model~\footnote{\revise{We also tested DeepSeek-V3.2 and GLM-4.7. The summaries generated by the three models have an average semantic similarity of 0.92, indicating that the teacher model has little impact on this step.}} to generate aspect-preference summaries, guiding the learning of open-source LLMs. 
With these summaries, we optimize the summarizer by SFT~\cite{SFT}.

\noindent$\bullet\quad$\textbf{Offline Summarization.} 
Given a summarizer, we can store the generated aspect-preference summaries offline for all the historical interactions between users and items.  
This step eliminates the need for repeated preference distillation, thereby reducing computational overhead and enhancing inference efficiency in the real-world deployment.

\subsection{Reasoner for Preference Matching}

Given all the aspect-preference summaries of a user and an item, Reason4Rec adopts a reasoner to evaluate their matching degree, generating reasons for why the user might like or dislike the item. 

\noindent$\bullet\quad$\textbf{Reasoner Instantiation.} 
Formally, given a user-item pair $(u, i)$, we aggregate the aspect-preference summaries from their historical interactions $\mathcal{H}_u$ and $\mathcal{H}_i$ in chronological order into a single prompt, tasking the reasoner to measure the preference matching between $u$ and $i$. 
The reasoner prompt is given as an example.
\begin{center}
\fcolorbox{black}{gray!6}{\parbox{0.98\linewidth}{
\centering 
\textbf{Reasoner Prompt} \\[2pt]
\raggedright 
\textbf{\#\#\# User Review History \#\#\#} 
$\langle$\textit{$\mathcal{H}_u$ organized as below}$\rangle$ \\
1. Title of Item 1\\
Positive Aspects: [Aspect 1], [Aspect 2], ... \\
Negative Aspects: [Aspect 1], [Aspect 2], ... \\
User Preference Elements: [Preference 1], [Preference 2], ... \\
... \\[3pt]
\textbf{\#\#\# Item Review History by Other Users \#\#\#} \\
$\langle$\textit{$\mathcal{H}_i$ organized in the same format as above}$\rangle$ \\[3pt]
\textbf{Task:} Analyze whether the user will like the new Music $i$ based on the user's preferences and the item's features. Provide your rationale in one concise paragraph.
}}
\end{center}

\noindent$\bullet\quad$\textbf{Reasoner Fine-tuning.} 
Given a historical interaction between user $u$ and item $i$, we propose using target item review $c_{ui}$ to guide the reasoner learning, because users would explicitly explain their preferences in verbalized reviews. 
However, directly using reviews for supervision suffers from noise issues, since user reviews often contain low-quality comments or preference-irrelevant details (see evidence in Section~\ref{sec:impact_of_verbalized}). 
Thus, we employ a generation-then-filter strategy: using ChatGPT first generates diverse reasons why user $u$ might like or dislike item $i$, and a reward model then selects the high-quality reasons based on target item review $c_{ui}$. The selected high-quality reasons are used for reasoner learning through SFT. 

\noindent\textbf{Reason Generation.}
As outlined in Algorithm~\ref{alg:reason_sampling}, given a user-item pair, we repeatedly sample candidate reasons from ChatGPT based on the reasoner prompt and evaluate reasons using the reward model until either a high-quality reason is obtained or the maximum iteration count is reached~\footnote{\revise{We set the maximum iteration count to 2, as a higher count does not significantly increase the yield of high-quality reasons.}}. 
Particularly, to avoid repeatedly generating reasons that contradict the user's binary preference, we incorporate the user's actual rating as a hint, except for the initial generations. 
This is implemented by appending the following hint statement to the reasoner prompt:
\begin{center}
\fcolorbox{black}{gray!6}{\parbox{0.98\linewidth}{
\textbf{Hint:} The user actually rated the item \textbf{$r_{ui}$} stars. The star ranges from 1 to 5, with 5 being the best. Use the hint but don't mention the user's rating in your response.
}}
\end{center}





\noindent\textbf{Reason Filter.} 
We introduce a reward model to filter candidate reasons based on their alignment with the target item review $c_{ui}$. 
Considering the alignment, the reward model assigns a score of 1 to high-quality reasons and 0 to low-quality reasons. 
Using this reward model, we can iteratively generate high-quality reasons using ChatGPT for the reasoner fine-tuning.
The detailed training and inference of the reward model can be found in Section~\ref{sec:reward}. 

\subsection{Predictor for Feedback Prediction} \label{sec:feedback_prediction}

Reason4Rec finally introduces a predictor for user feedback prediction.  
In this work, we investigate the classic rating prediction task~\cite{deepconn}.
The input for the predictor is the reasoning process generated by the summarizer and reasoner. 
In addition, considering that the user's historical ratings serve as a valuable reference for rating prediction~\cite{MF}, we also include ratings as input. 
The user's actual rating for the target item $r_{ui}$ is used for supervision. 

\noindent$\bullet\quad$\textbf{Predictor Instantiation.} 
Given a user-item pair $(u, i)$ with historical interactions $(\mathcal{H}_u, \mathcal{H}_i)$, we first aggregate the aspect-preference summaries from $\mathcal{H}_u$ and $\mathcal{H}_i$ in the same format as in the reasoner prompt, supplemented with historical ratings (see the following predictor prompt). 
Next, following previous work~\cite{expert}, we include the average ratings from all historical interactions in the prompt. 
Finally, the reason generated by the reasoner is appended to the \textit{Reason Placeholder}. 
The final prompt is shown below to predict the user ratings. 
To implement the predictor, we align the predicted scores with the ground-truth rating $r_{ui}$ by SFT.

\begin{center}
\fcolorbox{black}{gray!6}{\parbox{0.98\linewidth}{
\centering 
\textbf{Predictor Prompt} \\[2pt]
\raggedright 
\textbf{\#\#\# User Review History \#\#\#} \\
$\langle$\textit{$\mathcal{H}_u$ is organized in the same format as in the reasoner prompt, except that historical ratings are additionally included.}$\rangle$ \\[3pt]
1. Title of Item 1, {\textbf{Rating:} 5.0} \\
Positive Aspects: [Aspect 1], [Aspect 2], ... \\
Negative Aspects: [Aspect 1], [Aspect 2], ... \\
User Preference Elements: [Preference 1], [Preference 2], ... \\
... \\[3pt]
\textbf{\#\#\# Item Review History by Other Users \#\#\#} \\
$\langle$\textit{$\mathcal{H}_i$ organized in the same format as above}$\rangle$ \\[3pt]
{\textbf{\#\#\# Average Past Ratings \#\#\#}} \\
User's Average Rating (all previous ratings): 4.5 \\
Item's Average Rating (all ratings by other users): 3.7 \\[3pt]
{\textbf{\#\#\# Personalized Recommendation Analysis \#\#\#}} \\
$\langle$\textit{This section is referred to as \textbf{Reason Placeholder}, where we place the generated reason here.}$\rangle$ \\[3pt]
\textbf{Task:} Based on the above information, please predict the user's rating for $i$, (1 being the lowest and 5 being highest, directly give the rating without other content.) \\
\textbf{[Output Format]} Predicted Rating:\texttt{[Rating between 1 and 5]}
}}
\end{center}

\noindent$\bullet\quad$\textbf{Logit-weighted Decoding.} We adopt logit-weighted decoding for the final prediction, which transforms the predicted rating scores from discrete integer values to continuous scores---a method proven to offer greater precision and adopted in previous work~\cite{expert}.

Specifically, when decoding the rating token for a prediction task on a scale of 1 to N,  the LLM first obtains the logit values for each possible rating token (\eg the logits for token ``1'', ``2'', \dots, ``N''), denoted as \( l_1, l_2, \dots, l_N \).
These logits are then normalized into probabilities \( p_k \) for each rating \( k \) using the softmax function: $p_k = \frac{\exp(l_k)}{\sum_{j=1}^N \exp(l_j)}, k \in \{1, 2, \dots, N\}$. 
The final prediction is calculated as the expected value: $r_{ui} = \sum_{k=1}^N k \cdot p_k$, where each possible rating is weighted by its corresponding probability.

\noindent$\bullet\quad$\textbf{Efficient Rating Prediction.} 
As the predictor prompt instructs the LLM to produce a fixed output format that begins with "Predicted Rating:", this prefix remains constant across all predictions. Consequently, we can bypass decoding for these fixed tokens and focus solely on generating the final rating through logit‑weighted decoding. Restricting decoding to the single rating token can significantly reduce inference time and computational cost.

\subsection{Reward Model for Reason Judgment} \label{sec:reward}

We use a reward model to select high-quality reasons for the reasoner fine-tuning. 
Given candidate reasons and the corresponding target item reviews, the reward model assigns a score of 1 if the reason is high-quality and aligns with the target item review; otherwise, it assigns 0. 

\noindent$\bullet\quad$\textbf{Reward Model Training.} 
We fine-tune another QLoRA on the LLM as the reward model, which is trained to predict the user ratings based on the user review. 
The input is the predictor prompt template, where the \textit{Reason Placeholder} is filled by the target item review rather than the generated reasons. The reward model is trained to extract the user's true preferences from the target item review to enhance prediction accuracy. 

\textbf{Separate Training Data.} To prevent the reward model from predicting ratings solely by memorizing training data rather than capturing the user's true preferences from target item reviews, we separate the training data for the reward model and the reasoner. 
This separation ensures that the reward model is evaluated on unseen data, enabling a more reliable assessment of reason quality.

\begin{algorithm}[t]
    \caption{Generation-then-filter strategy}
    \label{alg:reason_sampling}
    \renewcommand{\algorithmicrequire}{\textbf{Input:}}
    \renewcommand{\algorithmicensure}{\textbf{Output:}}
    \begin{algorithmic}[1]
    \Require A teacher model LLM, a reward model $R$, the reasoner prompt $P$, an iteration count $T$, a Hint with user's actual rating.

    \State Initialize a high-quality reason set $\mathcal{S}_{\text{high}} \leftarrow \emptyset$;
    \State Initialize a reason $s \leftarrow \text{LLM}(P)$;

    \ForAll{$t \in \{1, \dots, T\}$}
        \State Get the reason quality $\text{Score} \leftarrow R(s)$;
        \If{$\text{Score} = 1$}
            \State Save $\mathcal{S}_{\text{high}} \leftarrow s$;
            \State \textbf{break};
        \Else
            \State Regenerate a new reason with hint: $s \leftarrow \text{LLM}(P, \text{Hint})$;
        \EndIf
    \EndFor

    \Ensure the high-quality reason in $\mathcal{S}_{\text{high}}$.
    \end{algorithmic}
\end{algorithm}

\noindent$\bullet\quad$\textbf{Reason Judgment.} 
Since the reward model is trained to capture the user's true preferences from the target item's review, a reason that better aligns with the user's review can achieve superior prediction accuracy, as evidenced in Section~\ref{sec:reasonable_of_reward}. 
Based on this insight, we define an evaluation score as:
\[
s_{\text{eval}} = \big| r_{ui} - r_{\text{reason}} \big| - \big| r_{ui} - r_{\text{review}} \big|,
\]
where $r_{ui}$ represents the ground-truth rating for user $u$ and item $i$. $r_{\text{review}}$ and $r_{\text{reason}}$ denote the predicted ratings by the reward model based on the target item review and a generated reason. And $s_{\text{eval}}$ measures the prediction difference: 
\begin{itemize}[leftmargin=*]
    \item If $s_{\text{eval}} < 0$, it suggests that the reason surpasses the raw review in rating prediction performance. 
    \item If $s_{\text{eval}} > 0$, a smaller $s_{\text{eval}}$ indicates a closer performance between the reason and review, suggesting superior alignment. 
\end{itemize}
By introducing a threshold $\tau$ \((\tau > 0)\), we identify all reasons of $s_{\text{eval}} < \tau$ as high-quality ones with the evaluation score as 1. 
Treating $\tau$ as a hyperparameter can balance the trade-off between the quantity and quality of the reasons.

\subsection{Overall Workflow of Reason4Rec}

In summary, Reason4Rec adopts a multi-expert collaborative framework to perform multi-step reasoning before recommendation, as illustrated in Figure~\ref{fig:framework}. We detail the overall workflow, including both the training and inference phases, as follows:

\noindent$\bullet\quad$\textbf{Training Phase.} 
In practice, we instantiate each expert as a separate QLoRA adapter~\cite{Qlora} on the same backbone LLM and train them sequentially with different supervision sources.  
First, we train the Summarizer to produce high-quality aspect–preference summaries using examples generated by the teacher model.  
Next, we use the target item review as supervision to filter high‑quality reasons that are consistent with the user’s preferences.  
We then use these filtered reasons to train the Reasoner to accurately analyze user preferences given their historical summaries.  
Subsequently, we train the Predictor to leverage the reasons generated by the reasoner to improve rating predictions, achieved by aligning the Predictor’s output ratings with the user’s ground‑truth ratings for the target items.

\noindent$\bullet\quad$\textbf{Inference Phase.} 
At inference time, we first use the Summarizer to generate aspect–preference summaries for the target user and item.
These summaries are then passed to the Reasoner, which generates reasons analyzing the user’s potential preference for the target item.
Finally, the Predictor takes the generated reasons together with the historical rating statistics and outputs the final rating prediction.

\section{EXPERIMENT}

In this section, we conduct a series of experiments to answer the following research questions: 
\begin{itemize}[leftmargin=*]
    \item \textbf{RQ1:} 
    How does Reason4Rec perform compared to other baseline methods in terms of accuracy in predicting user feedback (\ie ratings) and the quality of generated reasons?
    
    \item \textbf{RQ2:} Why is the multi-step reasoning strategy in Reason4Rec essential, and how does it compare in effectiveness to one-step reasoning strategies?
    
    \item \textbf{RQ3:} 
    How effective is our approach (the reason generation method and proposed reward model) in incorporating user-verbalized preference feedback?
\end{itemize}
    
    

\subsection{Experimental Settings}
\subsubsection{\textbf{Datasets.}} \label{sec:dataset}
We evaluate Reason4Rec on the Digital Music and Book categories from the Amazon Product dataset\footnote{\url{https://cseweb.ucsd.edu/~jmcauley/datasets/amazon/links.html}.}, and the Yelp Open dataset\footnote{\url{https://business.yelp.com/data/resources/open-dataset/}.}.
We use the entire Music dataset (64,706 interactions with 5,541 users and 3,568 items) for experiments, while for the Book and Yelp datasets, we adopt subsets due to the limitations of time and computational resources. For the Book dataset, we use data from the last two months (540,074 interactions with 174,050 users and 127,821 items), and for the Yelp dataset, we use data from the last six months (302,558 interactions with 173,099 users and 67,854 items).
For each dataset, we split it into training, validation, and test sets following a ratio of 8:1:1 based on the timestamps of interactions, ensuring that test interactions occur after all training and validation interactions to prevent information leakage~\cite{DBLP:journals/tois/JiS0L23}.
Regarding data filtering, following prior work~\cite{liu2019nrpa}, we adopt a 5-core setting.

\subsubsection{\textbf{Evaluation Metrics.}} 
To evaluate the accuracy of rating prediction, we use two standard metrics~\cite{expert,chatgpt_CIKM}: Mean Absolute Error (MAE) and Root Mean Square Error (RMSE). To evaluate reasoning quality, we employ BLEURT~\cite{Bleurt} and GPTScore~\cite{GPTScore} to measure the semantic alignment between the generated reasoning and the target item review. BLEURT\footnote{\url{https://github.com/google-research/bleurt}.} is an evaluation metric that leverages contextual embeddings to assess the semantic similarity between text pairs, providing a fine-grained measure of alignment. 
GPTScore, computed based on GPT-4o\footnote{\url{https://chatgpt.com/?model=gpt-4o}.}, analyzes the semantic alignment of the given text pairs and assigns a score between 0 and 100, with higher scores indicating better alignment. These metrics are commonly adopted in explainable recommendation tasks~\cite{XRec}.


\subsubsection{\textbf{Baselines.}}

\begin{table*}[t]
\setlength{\tabcolsep}{5mm}
\small
\centering
\caption{
Comparison of user feedback prediction performance between our Reason4Rec method and the baselines across the three evaluation datasets. 
The best results are highlighted in bold, and the second-best results are underlined.
}
\begin{tabular}{llcccccc}
\toprule
\multirow{2}{*}{\bf Type} & \multirow{2}{*}{\textbf{Method}} 
& \multicolumn{2}{c}{\textbf{Music}} 
& \multicolumn{2}{c}{\textbf{Book}} 
& \multicolumn{2}{c}{\textbf{Yelp}} \\
\cmidrule(lr){3-4} \cmidrule(lr){5-6} \cmidrule(lr){7-8}
& & \textbf{MAE} $\downarrow$ & \textbf{RMSE} $\downarrow$
& \textbf{MAE} $\downarrow$ & \textbf{RMSE} $\downarrow$
& \textbf{MAE} $\downarrow$ & \textbf{RMSE} $\downarrow$  \\
\midrule
\multirow{1}{*}{CF-based} 
& MF 
& 0.6188 & 0.8142 
& 0.6277 & 0.8565
& 0.7980 & 1.0711 \\
\midrule
\multirow{3}{*}{\shortstack[l]{Review-based}} 
& DeepCoNN 
& 0.6034 & 0.8057
& 0.6221 & 0.8403
& 0.8312 & 1.0665 \\
& NARRE 
& 0.5799 & 0.7881 
& 0.6242 & 0.8435
& 0.8177 & 1.0785 \\
& DAML 
& 0.5703 & \underline{0.7848} 
& 0.6214 & \underline{0.8371}
& \underline{0.7964} & \underline{1.0405} \\
\midrule
\multirow{4}{*}{LLM-based} 
& GPT-4o
& 0.7438 & 1.1069 
& 0.7591 & 1.1558
& 0.8766 & 1.3005 \\
& Deepseek-R1
& \textbf{0.5197} & 0.8849
& 0.6155 & 1.0019
& 0.8253 & 1.2959 \\
& Rec-SAVER 
& 0.6463 & 0.9262 
& 0.6645 & 0.9356
& 0.8295 & 1.1282 \\
& EXP3RT 
& 0.5608 & 0.8385 
& \underline{0.6135} & 0.9370
& 0.8306 & 1.2311 \\
\midrule
\multirow{1}{*}{ Ours} 
& \text{Reason4Rec}
& \underline{0.5442} & \textbf{0.7722} 
& \textbf{0.6029} & \textbf{0.8345}
& \textbf{0.7515} & \textbf{1.0343} \\
\bottomrule
\end{tabular}
\label{tab:feedback_acc}
\end{table*}

To comprehensively evaluate the proposed paradigm, Reason4Rec, we compare it against two categories of methods: review-base recommendation methods and LLM-based methods. Specifically, we select the following methods as baselines:

\begin{itemize}[leftmargin=*]
    \item \textbf{MF}~\cite{MF}: Matrix Factorization is a classical collaborative filtering method that predicts user ratings based solely on the historical rating matrix.

    \item \textbf{DeepCoNN}~\cite{deepconn}: 
    This method employs a Convolutional Neural Network (CNN) model to jointly learn item properties and user behaviors from review text to assist in rating prediction.

    \item \textbf{NARRE}~\cite{NARRE}: 
    This method employs an attention mechanism to prioritize the most useful reviews, enabling better extraction of user and item features to enhance recommendations.

    \item \textbf{DAML}~\cite{DAML}: This method models the interaction between user and item review documents to improve the representations of both users and items.

    \item \textbf{GPT-4o}~\cite{chatgpt_CIKM}: This approach uses GPT-4o to directly predict ratings based on the user's rating history. We adopted the same few-shot prompting template as \cite{chatgpt_CIKM}.

    \item \textbf{Deepseek-R1}~\cite{deepseek_r1}: This method leverages the native thinking mode of Deepseek-R1 to perform explicit reasoning before making rating predictions based on user's rating history. We adopt the same rating prediction prompt as Reason4Rec to ensure fair comparison.

    \item \textbf{Rec-SAVER}~\cite{ACL_findings}: This method supplies LLMs with the target item's metadata (e.g., title, categories, description) and the user's historical interactions (e.g., metadata of previously interacted items, user ratings, and raw reviews), and then instructs the LLMs to generate their reasoning process before providing the prediction.
    

    \item \textbf{EXP3RT}~\cite{expert}: 
    This method first constructs user and item profiles from historical reviews and ratings using LLMs. These profiles are then leveraged to reason about user preferences and predict potential ratings in a single step.
    
\end{itemize}

\subsubsection{\textbf{Implementation Details.}}
For review-based methods, we adopt the implementation\footnote{\url{https://github.com/ShomyLiu/Neu-Review-Rec}.} from previous work~\cite{liu2019nrpa}.
For all finetuned LLMRec methods (\ie Rec-SAVER, EXP3RT and our Reason4Rec), we use GPT-3.5-turbo\footnote{\url{https://platform.openai.com/docs/models\#gpt-3-5-turbo}.} as the teacher model and fine-tune the LLama3-8B model\footnote{\url{https://www.llama.com/docs/model-cards-and-prompt-formats/meta-llama-3/}.} with QLoRA~\cite{Qlora}, \revise{which updates only a small set of low-rank adapter parameters (about 45M trainable parameters, \ie less than 1\% of the full 8B model) and reduces GPU memory overhead by more than 70\% compared with full-parameter fine-tuning.} 
To accelerate the training and inference process, we leverage the Unsloth\footnote{https://github.com/unslothai/unsloth.} framework. 
\revise{In practice, each QLoRA-based module takes about 5 hours to fine-tune on one NVIDIA A100 GPU.}
To reduce computational costs, we randomly sample 12,000 user-item pairs from the training data for LLM instruction data, while conventional methods use the full train dataset. This sample size is sufficient to achieve strong performance, as demonstrated in our main results. All finetuned LLMRec methods are trained on the same 12,000 user-item pairs. For training the reward model, we use an additional 8,000 user-item pairs, ensuring no overlap with the instruct data. The hyper-parameter reward threshold ($\tau$) is set to 0.1, 0.2, and 0.08 for Music, Book, and Yelp, respectively. 
To address the context length limitations of LLMs, we follow previous work~\cite{bigRec} and select the ten most recent historical interactions for inclusion in the Reasoner and Predictor Prompts. We further analyze and discuss the impact of interactions volumes on model performance in Section~\ref{sec:review_volumes}. 
More details can be found in our code.

\subsection{Main Results (RQ1)}
We compare our Reason4Rec method with baseline methods on the accuracy of user feedback (rating) prediction (Predictor output) and the quality of generated reasons (Reasoner output).

\header{Accuracy of User Feedback Prediction}
The comparison of user feedback prediction performance is summarized in \tabref{tab:feedback_acc}, revealing two key findings:

\textbf{1)} Compared with traditional methods (\ie review-based and CF-based methods), our method achieves better performance on all three datasets, with an average MAE reduction of approximately 0.02 and an RMSE reduction of approximately 0.01. These superior results of Reason4Rec demonstrate that incorporating reasoning about user preferences and item features prior to generating the final prediction—i.e., applying deliberation—is effective in enhancing RecLLM performance, thereby validating the soundness of our formulation of deliberative recommendation;

\textbf{2)} Compared with other reasoning-enhanced methods (\ie LLM-based methods), our method substantially outperforms all competitors, achieving the best RMSE and MAE scores across all three datasets, except for the MAE on the Music dataset, which is lower than that of DeepSeek-R1. However, the RMSE of DeepSeek-R1 is notably higher—by more than 0.1—which highlights the advantage of our approach. This performance gap may be attributed to the fact that these reasoning-enhanced methods adopt a single-step reasoning process during training and lack alignment with users’ true preferences, in contrast to our superior multi-step reasoning with an expert-wise training strategy.

\begin{table}[tbh]
\setlength{\tabcolsep}{1.2mm}
\centering
\caption{Quality evaluation of generated reasons. ``GPT'' refers to ``GPTScore''. The best results are highlighted
in bold and second best results are underlined.}
\begin{tabular}{lcccccc}
\toprule
\multirow{2}{*}{\textbf{Method}} & \multicolumn{2}{c}{\textbf{Music}} & \multicolumn{2}{c}{\textbf{Book}} & \multicolumn{2}{c}{\textbf{Yelp}} \\
\cmidrule(lr){2-3} \cmidrule(lr){4-5} \cmidrule(lr){6-7}
& \textbf{GPT} & \textbf{BLEURT} & \textbf{GPT} & \textbf{BLEURT} & \textbf{GPT} & \textbf{BLEURT} \\
\midrule
DeepSeek-R1 & \underline{79.62} & \underline{0.3863} & \underline{76.24} & 0.4333 & \underline{71.98} & 0.4115\\
Rec-SAVER & 75.60 & 0.3652 & 72.45 & 0.4233 & 66.43 & 0.4102 \\
EXP3RT & 76.22 & 0.3840 & 73.60 & \underline{0.4373} & 64.28 & \underline{0.4275} \\
\textbf{Reason4Rec} & \textbf{80.53} & \textbf{0.4067} & \textbf{77.31} & \textbf{0.4731} & \textbf{72.70} & \textbf{0.4565} \\
\bottomrule
\end{tabular}
\label{tab:quality_eval}
\end{table}


\begin{table}[t]
\setlength{\tabcolsep}{0.6mm}
\centering
\caption{Human evaluation results on reason quality across three dimensions. The best results are highlighted in bold.}
\begin{tabular}{lccc}
\toprule
\textbf{Method} & \textbf{Reasonability} & \textbf{Persuasiveness} & \textbf{Preference Alignment} \\
\midrule
DeepSeek-R1 & 4.76 & 4.34 & 3.67 \\
\textbf{Reason4Rec} & \textbf{4.83} & \textbf{4.43} & \textbf{3.85} \\
\bottomrule
\end{tabular}
\label{tab:human_eval}
\end{table}

\header{Quality of Generated Reasons}
We next analyze the quality of the reasons generated by the Reasoner in Reason4Rec, in comparison with other reasoning-enhanced baselines. 
First, we automatically assess how well these reasons align with users' true preferences by measuring the semantic similarity between the generated reasons and the original user reviews. This evaluation is performed across all test sets, using GPTScore and BLEURT as the metrics.
As shown in Table~\ref{tab:quality_eval}, Reason4Rec demonstrates superior performance across both GPTScore and BLEURT metrics compared to all baseline methods, suggesting that the reasons generated by Reason4Rec exhibit stronger alignment with users' true preferences. This highlights the importance of leveraging verbalized user feedback and the effectiveness of the proposed expert-wise training strategies.

Moreover, to obtain a more comprehensive assessment of reason quality, we also conducted a human evaluation study following prior work~\cite{expert}. We randomly sampled 100 cases (50 each from Books and Music datasets) and employ three independent judges to assess the reasons generated by Reason4Rec and the strongest baseline (DeepSeek-R1). The evaluation was conducted using a 5-point scale (0-5) across three critical dimensions: \textbf{Reasonability}, which measures the logical coherence and plausibility of the generated reasons; \textbf{Persuasiveness}, which evaluates the effectiveness in convincing users about the recommendations; and \textbf{Preference Alignment}, which assesses the consistency with user's historical review preferences. The results of the human evaluation are presented in Table~\ref{tab:human_eval}. Both methods achieve high scores (above 4.0) in Reasonability and Persuasiveness, demonstrating the strong capability of LLMs in generating coherent and convincing reasons. Notably, Reason4Rec shows the most significant improvement over DeepSeek-R1 in Preference Alignment, with an increase of 0.18 points. This suggests the crucial role of verbalized user feedback as a supervisory signal for better aligning with user preferences.

\revise{
\header{Performance on Ranking Task}
To further evaluate whether Reason4Rec generalizes beyond rating prediction, we additionally assess its performance on ranking-based recommendation tasks. Specifically, for the Music and Book datasets, we construct a ranking test set by pairing each user's high-rated test items (\ie items with ratings $\geq 4$) with randomly sampled unseen items as negatives, resulting in a candidate set of 20 items for each user. We then rank the candidate items according to their predicted scores and evaluate the results using NDCG@5 and Precision@5. As shown in Table~\ref{tab:topk_perf}, Reason4Rec consistently achieves the best performance on both datasets and across both metrics. These results indicate that the proposed framework is not only effective for rating prediction, but also generalizes well to top-$K$ ranking tasks, demonstrating the broader applicability of deliberative reasoning in recommendation.
}

\begin{table}[t]
\setlength{\tabcolsep}{2mm}
\small
\centering
\caption{
\revise{Comparison of top-$5$ ranking performance of Reason4Rec and baseline methods. 
The best results are highlighted in bold, and the second-best results are underlined.}
}
\begin{tabular}{llcc}
\toprule
\textbf{Dataset} & \textbf{Method} & \textbf{NDCG@5} $\uparrow$ & \textbf{Precision@5} $\uparrow$ \\
\midrule
\multirow{9}{*}{\textbf{Music}}
& MF           & 0.2802 & 0.1701 \\
& DeepCoNN     & 0.2865 & 0.1782 \\
& NARRE        & 0.2905 & 0.1769 \\
& DAML         & \underline{0.3000} & 0.1782 \\
\cmidrule(lr){2-4}
& GPT-4o       & 0.2747 & 0.1633 \\
& Deepseek-R1  & 0.2885 & 0.1782 \\
& Rec-SAVER    & 0.2787 & 0.1707 \\
& EXP3RT       & 0.2925 & \underline{0.1796} \\
\cmidrule(lr){2-4}
& \text{Reason4Rec} & \textbf{0.3311} & \textbf{0.1980} \\
\midrule
\multirow{9}{*}{\textbf{Book}}
& MF           & 0.1629 & 0.0735 \\
& DeepCoNN     & 0.1889 & 0.0879 \\
& NARRE        & 0.1939 & 0.0894 \\
& DAML         & \underline{0.2197} & \underline{0.0971} \\
\cmidrule(lr){2-4}
& GPT-4o       & 0.1538 & 0.0690 \\
& Deepseek-R1  & 0.1822 & 0.0813 \\
& Rec-SAVER    & 0.1631 & 0.0714 \\
& EXP3RT       & 0.2118 & 0.0953 \\
\cmidrule(lr){2-4}
& \text{Reason4Rec} & \textbf{0.2365} & \textbf{0.1092} \\
\bottomrule
\end{tabular}
\label{tab:topk_perf}
\end{table}

\subsection{In-depth Analysis}

\subsubsection{\textbf{Impact of Multi-step Reasoning  (RQ2)}} \label{sec:each_step_important}
To evaluate the impact of multi-step reasoning in Reason4Rec, we first conduct an ablation study to analyze the necessity of the component in each step (\ie Summarizer, Reasoner and Predictor) and then assess the performance changes by replacing the multi-step reasoning mechanism with two alternative single-step reasoning strategies.

\header{Necessity of Each Expert}
An ablation study on Reason4Rec is conducted by removing each expert to assess its contribution. 
Specifically, we make the following changes in each experiment: 
1) \textbf{w/o Step 1} (\ie Summarizer): Directly using user historical reviews to replace the aspect-preference summaries as input for Reasoner and Predictor;  
2) \textbf{w/o Step 2} (\ie Reasoner): Remove the Reason Placeholder section from the Predictor Prompt and instruct the LLMs to directly predict the rating based solely on the aspect-preference summaries and historical ratings; 
3) \textbf{w/o Step 3} (\ie Predictor): Combine Reasoner and Predictor into a single model, asking the LLMs to simultaneously generate the match analysis and predicted ratings given the aspect-preference summaries and the user's historical ratings.

\begin{figure}[t]
  \centering
  \includegraphics[width=\linewidth]{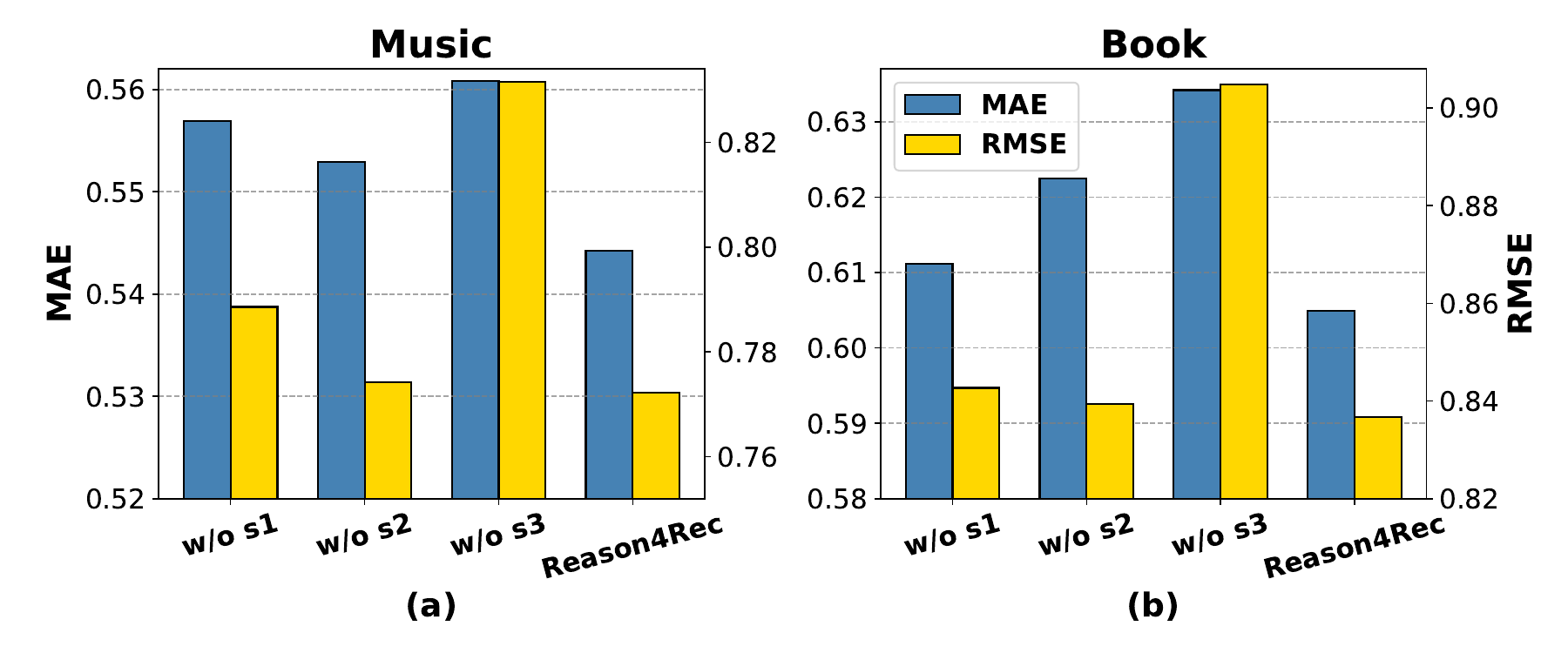}
  \caption{Necessity of Each Step. An ablation study on each step in Reason4Rec.}
  \label{fig:ablation}
\end{figure}

\begin{table}[t!]
\setlength{\tabcolsep}{1.5mm}
\centering
\caption{Reasoning quality associated with correct and
incorrect rating predictions}
\begin{tabular}{ccccccc}
\toprule
\multirow{2}{*}{\shortstack{\textbf{Correct}\\\textbf{Prediction}}} & \multicolumn{2}{c}{\textbf{Music}} & \multicolumn{2}{c}{\textbf{Book}} & \multicolumn{2}{c}{\textbf{Yelp}} \\
\cmidrule(lr){2-3} \cmidrule(lr){4-5} \cmidrule(lr){6-7}
& \textbf{GPT} & \textbf{BLEURT} & \textbf{GPT} & \textbf{BLEURT} & \textbf{GPT} & \textbf{BLEURT} \\
\midrule
Yes & \textbf{86.75} & \textbf{0.4206} & \textbf{84.28} & \textbf{0.4820} & \textbf{81.03} & \textbf{0.4568}\\
No & 79.20 & 0.4037 & 75.90 & 0.4713 & 71.71 & 0.4464 \\
\bottomrule
\end{tabular}
\label{tab:reason_quality}
\end{table}

The ablation results on Amazon Music and Book datasets are presented in \figref{fig:ablation}, from which we make the following observations: 
\begin{itemize}
    \item The removal of each expert results in an increase of the rating prediction errors on both Music and Book datasets, as reflected by higher MAE and RMSE values in the figure; 

    \item The MAE and RMSE increase significantly on both datasets after removing Step 3, underscoring the importance of separately training match analysis and rating prediction. One potential reason is that simultaneously training the LLM to complete the reasoning and rating tasks imposes significant challenges, resulting in degraded performance; 

    \item The removal of Step 2 leads to different impacts on RMSE and MAE metrics. The relatively small increase in RMSE can be explained by its heightened sensitivity to extreme prediction errors (e.g., predicting 1 for a true rating of 5). Even without reasoning, LLMs can still make roughly accurate rating predictions based on user reviews and historical data, thus avoiding extreme errors. In contrast, MAE treats all errors equally, making it better at detecting subtle prediction inaccuracies. The substantial increase in MAE therefore demonstrates that Reasoner plays a vital role in enabling LLMs to make more precise rating predictions.


\end{itemize}

\begin{figure}[t]
  \centering
  \includegraphics[width=\linewidth]{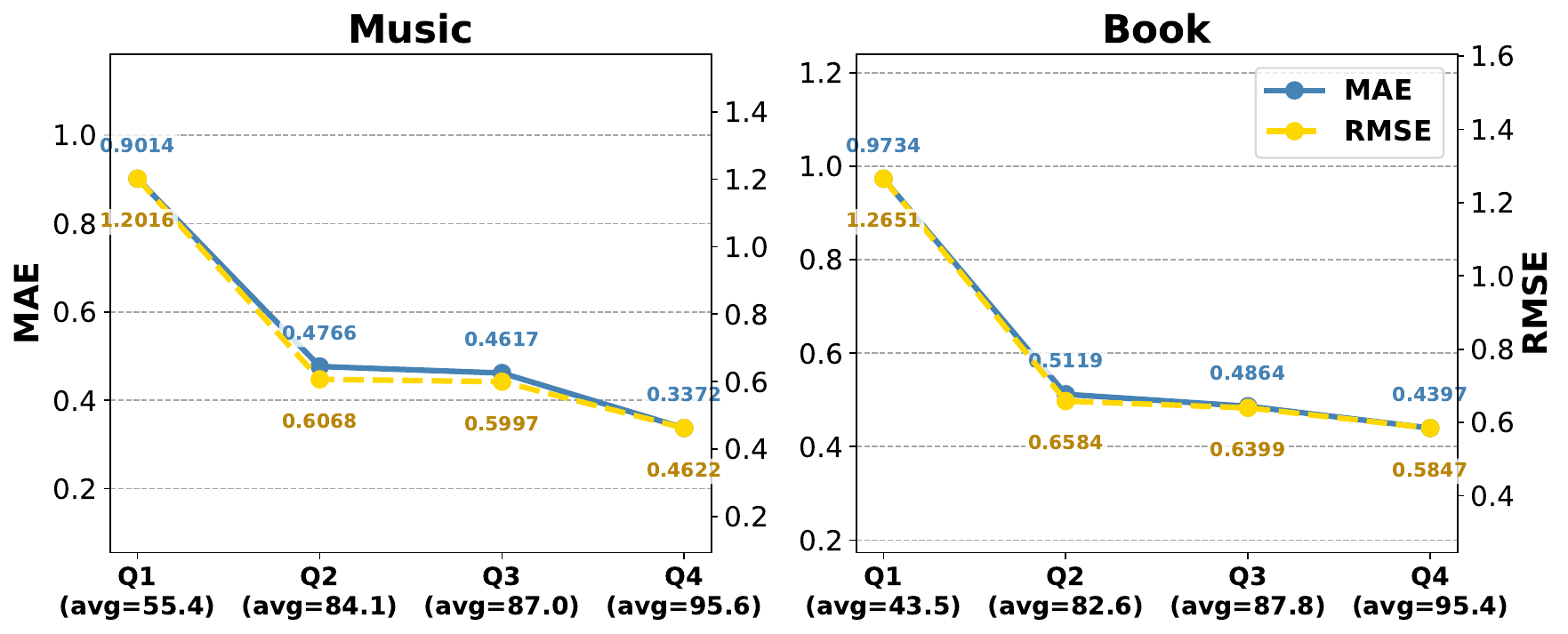}
  \caption{\revise{Prediction error across reasoning quality quartiles. Samples are grouped into four quartiles (Q1--Q4) by GPTScore. The value below each quartile label indicates the average GPTScore within that group.}}
  \label{fig:reasoning_quality}
\end{figure}

\header{Reasoning-Accuracy Relationship Analysis} \revise{We further investigate the relationship between reasoning quality and prediction accuracy from two complementary perspectives. As shown in Table~\ref{tab:reason_quality}, correct predictions (with error $<$ 0.01) consistently exhibit higher reasoning quality. Furthermore, as illustrated in Figure~\ref{fig:reasoning_quality}, test samples associated with higher reasoning quality consistently yield lower MAE and RMSE values. Taken together, these two perspectives demonstrate that high-quality reasoning facilitates accurate preference prediction, underscoring the importance of improving reasoning quality for reliable rating estimation.}

\revise{We also examine challenging cases in which high-quality reasoning still leads to prediction errors. In these cases, the reasoning correctly captures the user's preference direction but fails to accurately capture the rating magnitude, \eg predicting 5 stars when the ground-truth rating is 4. This suggests that while reasoning can recover preference-related signals, converting such signals into exact ratings requires additional calibration to each user's subjective rating scale. }

\begin{figure}[t]
  \centering
  \includegraphics[width=\linewidth]{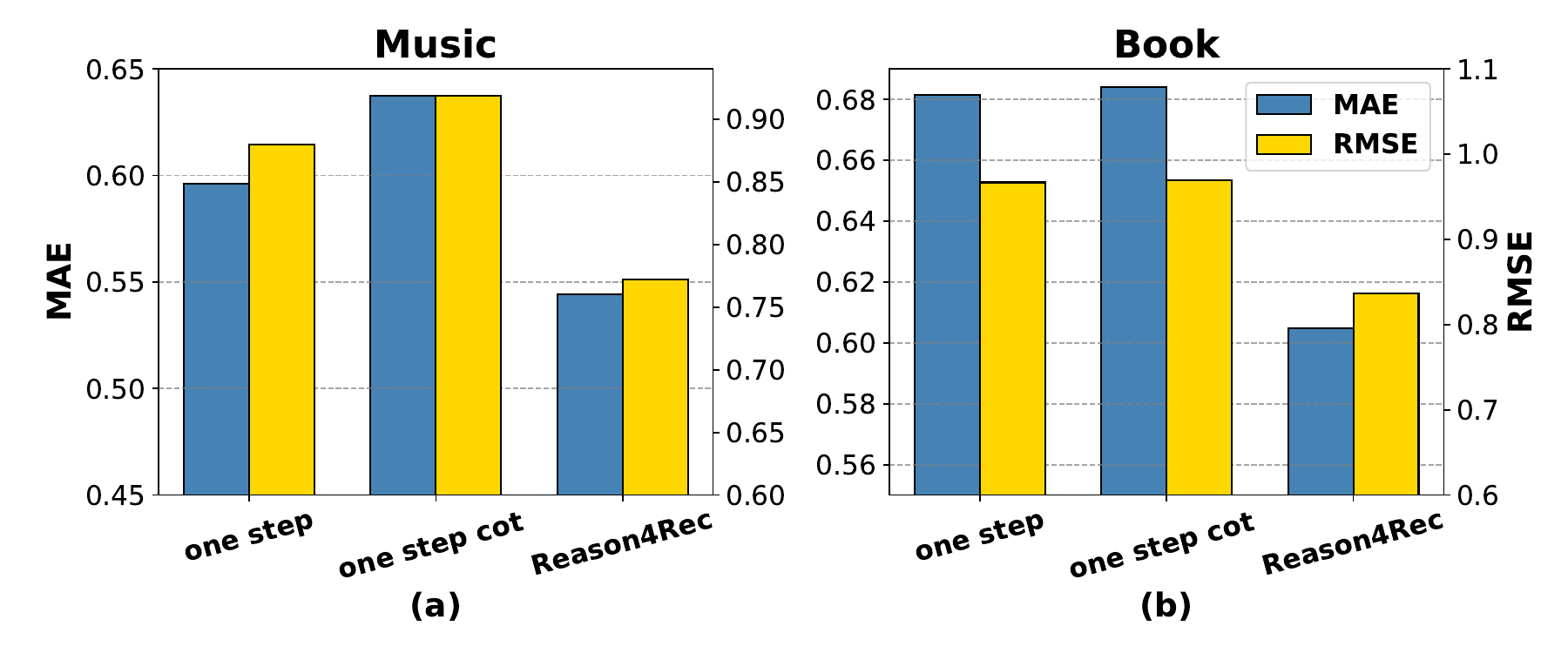}
  \caption{One-Step v.s. Multi-Step. Performance comparison between Reason4Rec's multi-step reasoning strategy and two alternative one-step reasoning strategies.}
  \label{fig:one_step}
\end{figure}

\begin{table}[b]
\centering
\caption{\revise{GPTScore of generated reasoning on training and test sets. $\downarrow$ denotes the quality drop from training to test set.}}
\label{tab:reasoning_quality_drop}
\begin{tabular}{lcccc}
\toprule
\multirow{2}{*}{} & \multicolumn{2}{c}{\textbf{Music}} & \multicolumn{2}{c}{\textbf{Book}} \\
\cmidrule(lr){2-3} \cmidrule(lr){4-5}
 & Train & Test & Train & Test \\
\midrule
One Step CoT & 83.48 & 74.45{\small($\downarrow$9.03)}  & 82.08 & 71.17{\small($\downarrow$10.91)} \\
Reason4Rec   & 83.97 & 80.53{\small($\downarrow$3.44)} & 82.87 & 77.31{\small($\downarrow$5.56)} \\
\bottomrule
\end{tabular}
\end{table}






\header{One-step v.s. Multi-step} 
To verify the effectiveness of the multi-step reasoning strategy in Reason4Rec, we compare it with two alternative strategies: (1) One-step reasoning, which generates both an entangled reasoning process and the ratings in a single step, and (2) Chain-of-Thought (CoT) reasoning, which generates the outputs of our three reasoning steps in one single prompt.
We prepare the datasets separately for the two strategies and train the LLMs accordingly. 
In \figref{fig:one_step},  we summarize the results and make the following observations: 
1) Both one-step reasoning and CoT reasoning strategies lead to a notable increase in prediction errors on both Music and Book datasets, as reflected by the higher MAE and RMSE values;
2) The performance of the CoT reasoning strategy is inferior to that of the simpler one-step reasoning strategy. 
These results indicate the complexity of multiple tasks in the CoT reasoning strategy is hard to learn in one step and underscore the importance of our multi-step strategy.

\revise{To further investigate why one-step CoT underperforms and validate the necessity of our multi-step reasoning design, we compare the reasoning quality (GPTScore) of both methods on training and test sets respectively. As shown in Table~\ref{tab:reasoning_quality_drop},  both methods achieve comparable reasoning quality on the training set. However, when generalizing to the test set, One-Step CoT suffers a 2$\times$ larger quality drop compared to Reason4Rec. We attribute this to the difficulty of learning complex multi-task reasoning in a single step, which causes the LLM to overfit to the training data rather than truly learning the underlying reasoning skills, thus leading to poor generalization. This validates the necessity of our multi-step design in reducing the learning complexity at each stage and improve reasoning generalization of reasoning quality.
}

\subsubsection{\textbf{Effect of Reason Generation Approaches (RQ3)}} \label{sec:impact_of_verbalized}
To validate the effectiveness of our reason generation approach for Reasoner, we compare it against four alternative approaches:
1) \textbf{Without reason}: Remove Reasoner from Reason4Rec entirely and instruct the LLMs to directly predict the rating based solely on the aspect-preference summaries and historical ratings, as done in the ablation study;
2) \textbf{Review-guided reasons}: Use the target item reviews as direct supervision for the LLMs' learning process in Reasoner;
3) \textbf{Review-inferred reasons}: Use ChatGPT to generate reasons inferred from the target item reviews and employ these generated reasons as supervision for Reasoner;
4) \textbf{History-inferred reason}: Use ChatGPT to generate reasons inferred from the aspect-preference summaries using the Reasoner Prompt and directly use them as supervision for Reasoner without applying filtering through our reward model.

\begin{table}[t]
\centering
\caption{Effect of different reason generation approaches.}
\begin{tabular}{lcccc}
\toprule
\multirow{2}{*}{\bf Method} & \multicolumn{2}{c}{\bf Music} & \multicolumn{2}{c}{\bf Book} \\ \cmidrule(lr){2-3} \cmidrule(lr){4-5}
                        & \bf MAE    $\downarrow$         & \bf RMSE   $\downarrow$        & \bf MAE $\downarrow$            & \bf RMSE    $\downarrow$       \\ \midrule

Without Reason          & \underline{0.5529}          & \underline{0.7742}          & \underline{0.6225}          & 0.8394          \\
\revise{Review-guided} Reason              & 0.5825          & 0.7812          & 0.6376          & 0.8390          \\
Review-inferred Reason   & 0.5723          & 0.7816          & 0.6306          & 0.8420          \\
History-inferred Reason        & 0.5640          & 0.7744          & 0.6305          & \underline{0.8376}         \\
\textbf{Reason4Rec}                    & \textbf{0.5442} & \textbf{0.7722} & \textbf{0.6029} & \textbf{0.8345} \\
\bottomrule
\end{tabular}
\label{tab:reason-gen}
\end{table}



In \tabref{tab:reason-gen}, we present the results and make the following observations:

1) All four alternative approaches underperform our proposed method on both Music and Book datasets, 
demonstrating the advantages of our approach in collecting quality reasons for LLM training;
2) Review-inferred Reason performs worse than Without Reason on both datasets.
This indicates that directly using target item reviews to generate reasons can have a negative impact. A possible reason is that our objective is to train LLMs to perform reasoning based on historical interactions \revise{(\ie both historical ratings and reviews)} without relying on target item reviews. Using reasons derived directly from target item reviews introduces the risk that the reasoning logic relies on preferences not reflected in the historical interactions.
Such inconsistencies disrupt the training process and  result in worse performance;

3) History-inferred Reason also performs worse than ``Without Reason'' on most metrics. 
This indicates that directly supervising with GPT-generated reasons without refinement is ineffective, underscoring the importance of using verbalized preference feedback as supervision.

\subsubsection{\textbf{Effectiveness of Reward Model (RQ3)}} \label{sec:reasonable_of_reward}

\begin{table}[t]
\centering
\caption{Rating prediction accuracy of reward model using different types of reasons.}
\begin{tabular}{lcccc}
\toprule
\multirow{2}{*}{} & \multicolumn{2}{c}{\textbf{Music}} & \multicolumn{2}{c}{\textbf{Book}} \\ \cmidrule(lr){2-3} \cmidrule(lr){4-5}
                                & MAE $\downarrow$            & RMSE $\downarrow$           & MAE $\downarrow$            & RMSE   $\downarrow$         \\ \midrule
Without Reason               & 0.6798          & 0.9359          & 0.6186          & 0.8298          \\
Review-guided Reason      & \textbf{0.294}   & \textbf{0.4527}  & \textbf{0.3329}  & \textbf{0.4597}  \\
Review-inferred Reason      & \underline{0.4238}           & \underline{0.568}           & \underline{0.4214}           & \underline{0.5394}           \\
Hint-inferred Reason        & 0.4879           & 0.6876          & 0.5174           & 0.7137           \\
History-inferred Reason     & 0.7257           & 1.071           & 0.6842           & 0.9686           \\ \midrule
\end{tabular}
\label{tab:reward_effect}
\end{table}

The reward model should be capable of assessing whether the generated reasons align with the ground-truth preferences (as reflected in the target item reviews) by utilizing these reasons for rating prediction.
We conduct experiments to verify whether the reward model successfully acquires this capability. Specifically, we construct different types of reasons with known similarity levels to target item reviews to compare the prediction accuracy. We include the four types of reviews defined in Section~\ref{sec:impact_of_verbalized}, along with an additional type: Hint-Inferred Reason. This type generates reasons using ChatGPT, based on aspect-preference summaries and our reasoner's prompt, with the user's actual rating included as a hint. Based on their similarity to ground-truth reviews, the types of reasons are ranked from highest to lowest as follows: Review-Guided, Review-Inferred, Hint-Inferred, and History-Inferred Reason. Table~\ref{tab:reward_effect} summarizes the reward model's prediction accuracy using each type of reason. The results show that the prediction accuracy ranking aligns with the known similarity ranking, verifying that the reward model effectively evaluates the quality of the generated reviews.

\subsubsection{\textbf{Impact of Reward Threshold}}
We investigate how the reward threshold $\tau$ affects model performance by conducting experiments with different threshold on the Yelp dataset. As shown in \tabref{tab:threshold}, the selection of $\tau$ involves a trade-off between reason quality and quantity. When $\tau$ is set too small (\ie 0.02), the limited training data leads to suboptimal performance. Conversely, larger $\tau$ values (\ie 0.10) yield more training samples but result in degraded performance due to compromised reason quality. 
\revise{
In practice, we select $\tau$ by performing a grid search on the validation set and choosing the value that yields the best downstream recommendation performance. The validation set is drawn from the same chronological data split used in our main experiments, \ie the 8:1:1 train/validation/test split described in Section~\ref{sec:dataset}. We first choose an initial $\tau$ that retains a moderate proportion of candidate reasons (around 50\%) as high-quality reasons, and then perform a local grid search around this value. For example, on the Yelp dataset, we use $\tau=0.06$ as the starting point and search over $\{0.02, 0.04, 0.06, 0.08, 0.10\}$. For new domains, we recommend the same retention-based selection strategy: first choosing an initial $\tau$ that retains approximately 50\% of candidate reasons, and then performing a local grid search around this value to determine the optimal threshold.}

\revise{
The optimal $\tau$ varies across datasets (\eg 0.08 for Yelp and 0.2 for Book). We attribute this mainly to differences in reasoning diversity across domains, which lead to different requirements on the amount of high-quality reasons needed for effective training. For example, domains with more diverse reasoning patterns, such as Book, require a larger $\tau$ to retain sufficient candidate reasons for training, while more focused domains, such as Yelp, allow for a stricter threshold to emphasize reason quality.
}

\subsubsection{\textbf{Impact of Interaction Volumes.}}
\label{sec:review_volumes}
\begin{figure}[t]
  \centering
  \includegraphics[width=\linewidth]{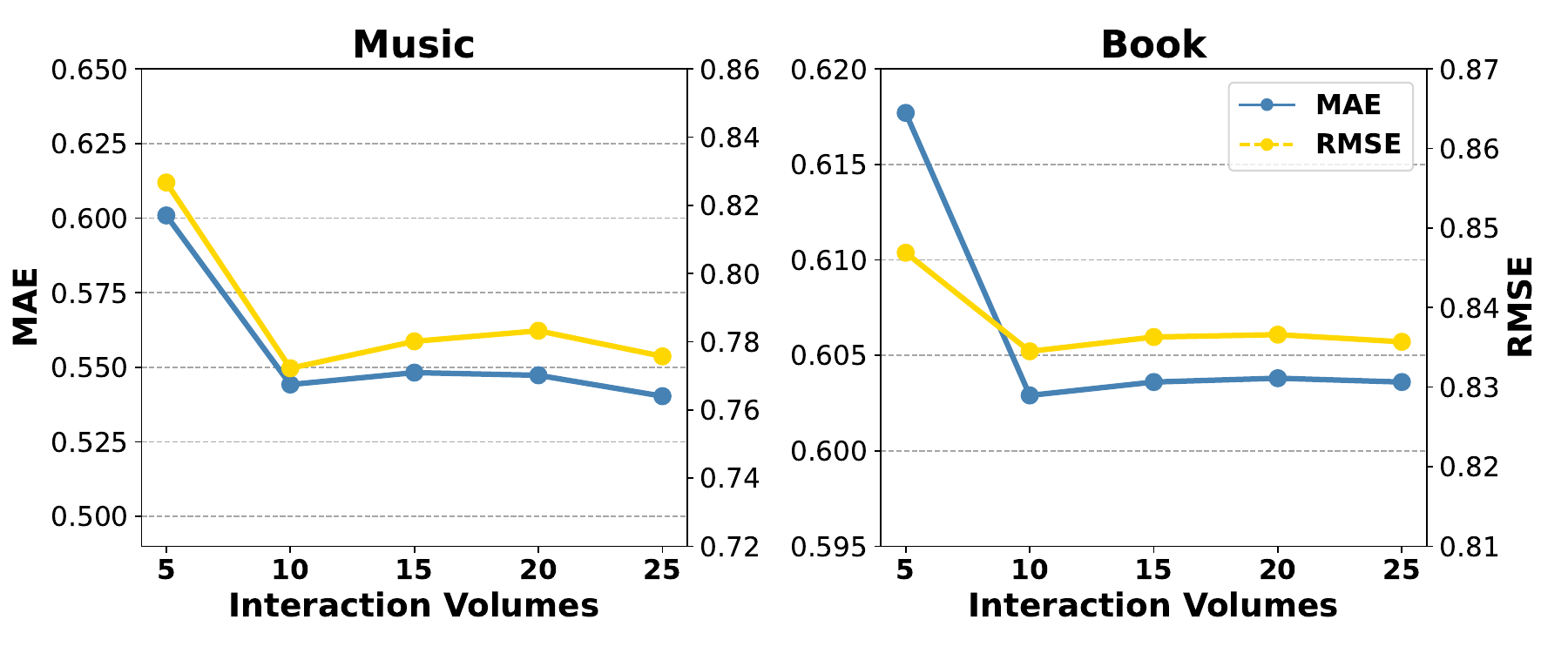}
  \caption{Prediction Accuracy v.s. Interaction Volumes.}
  \label{fig:review_volumn}
\end{figure}

We analyze how the number of historical interactions affects Reason4Rec's performance by including different numbers of most recent interactions from users' and items' histories. 
As illustrated in \figref{fig:review_volumn}, expanding the number of histories from five to ten significantly enhances prediction accuracy, as additional histories provide a better understanding of user preferences. However, beyond ten histories, the impact on prediction accuracy becomes negligible, with changes in MAE and RMSE being limited to 0.004 and 0.008, respectively. These results suggest that ten \revise{historical interactions} might be adequate to capture user preference on Music and Book datasets, possibly because some short-term interests dominate in these scenarios. 

\begin{table}[t]
\setlength{\tabcolsep}{5mm}
\centering
\caption{Impact of reward threshold $\tau$ on model performance.}
\label{tab:threshold}
\begin{tabular}{cccc}
\toprule
$\tau$ & RMSE & MAE & Training Data Size \\
\midrule
0.02 & 1.0908 & 0.7751 & 2,705 \\
0.04 & 1.0418 & 0.7586 & 5,315 \\
0.06 & 1.0444 & 0.7553 & 5,964 \\
0.08 & \textbf{1.0343} & \textbf{0.7515} & 6,587 \\
0.10 & 1.0505 & 0.7701 & 7,158 \\
\bottomrule
\end{tabular}
\end{table}

\subsubsection{\textbf{Inference Cost Analysis}}
To evaluate the inference efficiency of Reason4Rec, we compare its inference time and generation cost with other baselines with reasoning. As shown in Table~\ref{tab:overhead-analysis}, Reason4Rec achieves comparable average inference time and generates a similar number of tokens as Rec-SAVER and EXP3RT, while generating significantly fewer tokens than Deepseek-R1. Specifically, it requires approximately 6 seconds and generates an average of 148 tokens per prediction. This efficiency is primarily attributed to two factors: 1) the Summarizer is preprocessed in advance, as discussed in Section~\ref{sec:step1}, leaving only the last two steps to be executed during inference; and 2) the Predictor only decodes a single token during the final step, as explained in Section~\ref{sec:feedback_prediction}. These designs effectively minimize computation costs.

\begin{table}[t]
\centering
\caption{
Inference cost analysis of the average inference time and the average number of tokens generated per prediction. Experiments were conducted on an NVIDIA A800 GPU using 100 data points from the Book dataset.
}
\begin{tabular}{lcc}
\toprule
Method & Avg. Inference Time (s) & Avg. Tokens Generated \\ \midrule
Reason4Rec   & 5.86           & 147.78           \\
Rec-SAVER    & 6.43           & 175.59           \\
EXP3RT & 5.62           & 150.74           \\ 
Deepseek-R1  & /\ & 711.69 \\ \bottomrule
\end{tabular}
\label{tab:overhead-analysis}
\end{table}

\subsubsection{\textbf{Case Study}}
\begin{figure}[t]
  \centering
\includegraphics[width=\linewidth]{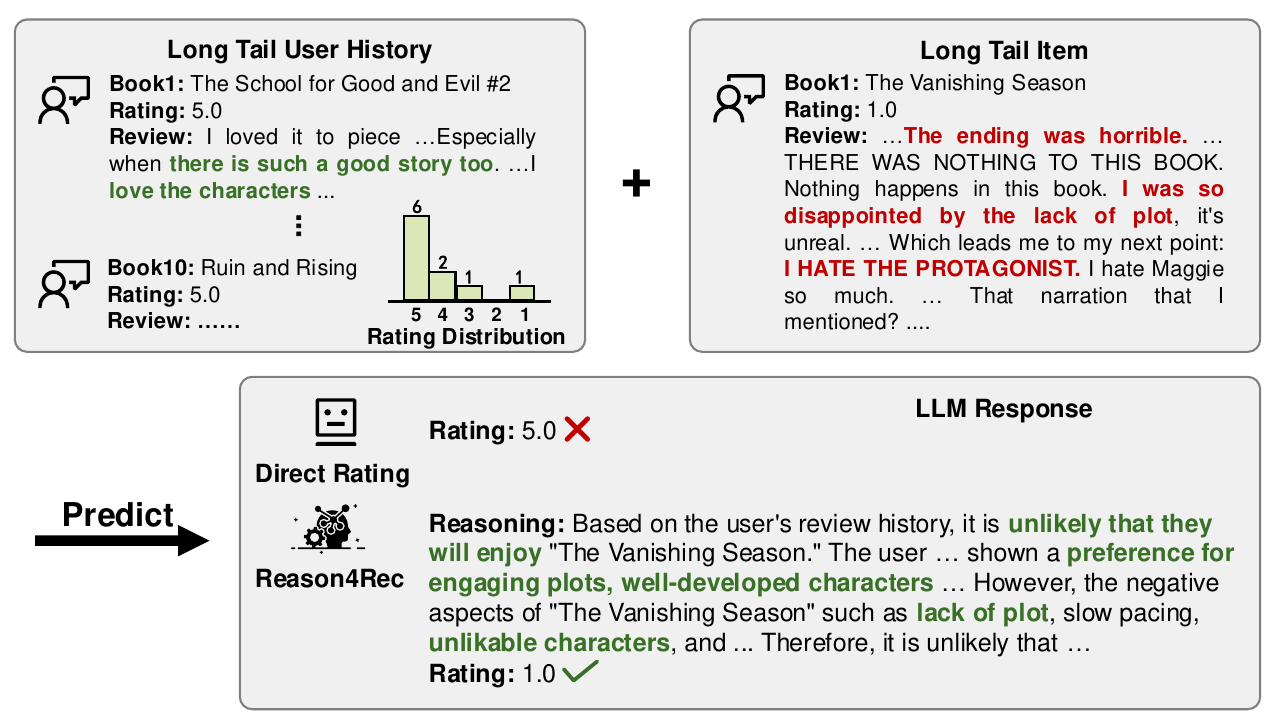}
  \caption{Case study on cold-start scenarios.}
  \label{fig:case}
\end{figure}

We conduct a case study on long-tail scenarios to intuitively explain how reasoning enhances prediction, as illustrated in Figure~\ref{fig:case}. In this case, the user exhibits a long-tail rating behavior, predominantly giving 5-star ratings, while the target item has only a single historical interaction. If we instruct the RecLLM to directly predict the rating, it tends to depend on the user's past rating patterns, overlooking the specific preferences expressed in their review contents, which leads to inaccurate predictions. In contrast, Reason4Rec extracts both user preferences and item features, and identifies the mismatch between them through reasoning, thereby accurately predicting user dislike of the target item. 
This demonstrates the effectiveness of reasoning in handling cold-start scenarios through comprehensive preference analysis.

\section{CONCLUSION AND FUTURE WORK}
In this work, we addressed the limitations of existing RecLLMs in complex scenarios by introducing a novel Deliberative Recommendation task, which emphasizes explicit reasoning before predicting user feedback. 
To achieve this, we proposed the Reason4Rec framework, which enables multi-step reasoning via three collaborative experts with three core reasoning capabilities: Preference Distillation, Preference Matching, and Feedback Prediction. 
We aligned the reasoning process with users' true preference by using verbalized user feedback, \ie reviews. 
Through extensive experiments on three real-world datasets, Reason4Rec demonstrated superior performance in both prediction accuracy and reasoning quality, highlighting the significance of slow thinking in recommendation tasks.

Despite its promising results, Reason4Rec is an initial attempt on Deliberative Recommendation, leaving many future directions. 
First, we only adopt user reviews as verbalized user feedback. \revise{A promising future direction is to enrich the reasoning context from two perspectives. On one hand, future work can explore more adaptive history selection mechanisms, such as attention-based context selection, to better prioritize the most relevant past interactions for deliberative reasoning.
On the other hand, future work can also consider modeling the user-item interaction history as a graph~\cite{liang2025concrete}. By learning from relational correlations and periodic events~\cite{liang2023learn} and using structure augmentation to handle islanded nodes~\cite{li2025eyes}, we can extract richer and more structured feedback signals, enabling the Reasoner to reason more accurately about user preferences.
Second, although Reason4Rec achieves comparable or even lower inference costs than Rec-SAVER and EXP3RT, there is still room to improve its efficiency. Future work could explore designing LLM acceleration algorithms tailored to Deliberative Recommendation.
Third, we only treat Reason4Rec as a recommendation model for predicting user feedback. Given that the reasons generated by Reason4Rec can explicitly express user preferences, future work could consider leveraging such explicit reasoning to construct more interpretable user simulators~\cite{zhang2025llm}.
Fourth, it is promising to deploy Reason4Rec on real-world recommendation platforms, such as OpenP5~\cite{xu2024openp5}, to better explore how to combine LLM-based deliberative reasoning with practical applications, enabling more systematic evaluation and broader adoption of deliberative recommendation in diverse real-world scenarios.}


\small
\bibliographystyle{IEEEtran}
\bibliography{main}

\end{document}